\documentclass[prb,preprint,showpacs,floats]{revtex4}

\usepackage{graphics}
\usepackage{epsfig}
\usepackage{dcolumn}
\usepackage{amsmath}

\begin{document}
\title{ Terahertz-field induced tunneling current with non-linear effects 
in a double quantum well coupled to a continuum }

\author{Marcelo Z. Maialle$^{1,2}$, Marcos H. Degani$^{1,3}$,
Justino R. Madureira$^4$, and Paulo F. Farinas$^{1,5,6}$}

\affiliation{$^1$DISSE - Instituto Nacional de Ci\^ encia e Tecnologia de Dispositivos
Semicondutores}
\affiliation{$^2$Liceu Vivere, R. Duque de Caxias Norte 550,
Pirassununga-SP, 13635-000, Brazil}\email{ mzmaialle@liceuvivere.com.br}
\affiliation{$^3$Haras Degani, 
Av. Fioravante Piovani 1000, 
Itatiba-SP, 13257-700, Brazil}
\affiliation{$^4$Universidade Federal de 
Uberl\^andia, Ituiutaba, MG, Brazil}
\affiliation{$^5$Departamento de F\' \i sica, Universidade Federal de S\~ ao Carlos,
13565-905, S\~ ao Carlos, SP, Brazil}
\affiliation{$^6$Instituto de F\' \i sica, Universidade Federal do Rio de Janeiro,
21645-970, Rio de Janeiro, RJ, Brazil}

\date{\today}

\begin{abstract}
We have theoretically investigated the tunneling current induced by a 
terahertz (THz) field applied to an asymmetric double quantum well. 
The excitation couples an initially localized state
to a nearby continuum of extended states. 
We have shown that the calculated current has similar features as those 
present in the optical spectra, such as interference effects due to 
the interaction between the continuum and the localized states,
in addition to many-photon transition effects.
The induced current is calculated as a function of the intensity
of the THz field. A second THz field
is used to yield non-linear processes, useful to control the interference 
effects. We believe that part of the issues studied here can be usefull for
the integration of novel switching mechanisms based on optics (THz) and 
electronic current.

\end{abstract}
\pacs{71.35Pq, 73.21.La, 78.55Cr, 78.67.Hc}

\maketitle

\section{Introduction}

Fast switching mechanisms are desirable for a number of applications, including
in quantum information processing. A limitation on the time-scale performance
of these mechanisms is usually a consequence of
the prevalence of incoherent processes,
such as the slow dynamics of population decay. 
One possible way to overcome this problem is to restrict the processes
to the faster quantum coherent dynamics, as in the case 
studied by Wu {\it et al.}\cite{wu} in which a scheme was proposed for an
all-optical switching mechanism based on the virtual excitation 
of excited states by two independent terahertz (THz) fields.

The system investigated by Wu {\it et al.}\cite{wu} is a double quantum well separated
by a thin barrier from a continuum of states (see Fig.~1). By exciting
this system with a THz (probe) field one sets the ground state (which is occupied
by electrons from a nearby reservoir) in resonance with two almost
degenerate excited states, both coupled to the continuum. This coupling
creates a Fano-like interference\cite{fano} which manifests itself as a strong 
suppression of the field absorption. As a result, the probe THz spectra show
two asymmetric peaks with an almost zero absorption 
in between the peaks.\cite{schmidt,faist}
When a second (switch) THz field is used, with a frequency that sets the
resonance between the pair of excited states to a further excited state,
there occurs a substantial modification of this interference. 
In this way, one can control the probe absorption with the switch field,\cite{harris} 
noticeably when the inference effect is intense. 

The novelty of the proposed all-optical switching device is in the fact that it
deals with virtual excitation of the highly energetic state, therefore
it can be quickly switched on and off. The calculations to support
the proposed model was done with a set of coupled
Maxwell-Schr\"odinger equations for the occupations of the quantum levels
and for the exciting THz fields.\cite{wu} The focus of the attention was on the
temporal evolution of the probe transmission under the control
of the switch field.

In the present work we obtain similar results looking at the 
THz-field--induced tunneling current. We have solved numerically
the time evolution of a one-electron state, initially
localized in the asymmetric double well, in the presence
of an applied THz field pulse (probe). Interference effects due to interaction 
with the continuum are observed in the tunneling current
as well as many-photon processes. A second field (switch) can be used
to control this interference, thus opening the possibility
to create switching mechanisms, which in the present case have
a signature in the tunneling current. This may be usefull for the integration
of optical and electronic schemes in a switching mechanism.

\section{Theory}

\begin{figure}[h]
\centerline{\epsfig{file=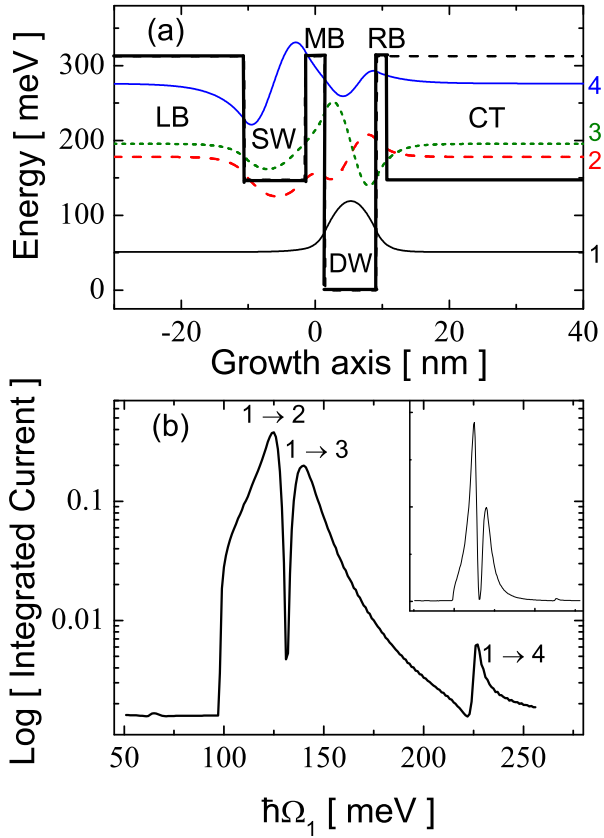,width=8.0cm}}
\caption{(Color online)
 (a) Potential profile of the double well system coupled
to the continuum.
The wavefunctions depicted serve to illustrate the processes
and are those of four  really bound states, taken from a system
with no continuum region to the right, that is, with the right barrier being
composed of Al$_{0.4}$Ga$_{0.6}$As, as illustrated by the dashed lines in the
right-side region.
(b) Log-scale graph of the integrated current induced 
by the probe field of intensity $F_1$=10 kV/cm
as function of the field frequency (energy).
The envelope of this oscillating field is a Gaussian of width 2 picoseconds. 
Labels refer to transitions between quantum states shown in (a).
The inset shows the same spectra in a linear vertical scale.}
\end{figure}

The system investigated is an asymmetric double quantum well 
which is separated from a continuum of states
by a thin potential barrier (RB), as depicted in Fig.~1.
An electron in the ground state, localized in the deeper quantum well (DW),
is excited by the probe field and may tunnel to the
continuum. The current associated with this process is calculated from
the time evolution of the single initially localized state
that the electron occupies at $t=0$.
This is done within the effective-mass approximation 
for an one-electron problem in the conduction band. 
A second (switch) oscillating field can be used to control this 
tunneling current.

The problem is solved numerically using the following 
one-dimensional hamiltonian
\begin{eqnarray}
H &=& - \frac{\hbar^2}{2m^*}\frac{d^2}{dz^2} + V(z)
-e z F_1(t) \sin \left( \Omega_1 t \right) \nonumber \\ 
&& -e z F_2(t) \sin \left(\Omega_2 (t-T) +\phi \right),
\label{H}
\end{eqnarray}
where $m^*$ is the electron effective mass, assumed uniform
throughout the system, $V(z)$ is the potential due to the
material band-offsets,
and $F_1(t)$ ($F_2$) is the envelope of the applied
probe (switch) electric field oscillating with frequency
$\Omega_1$ ($\Omega_2$) and polarized along the $Z$ direction.
 $\phi$ represents a possible relative
phase between the fields. The probe and switch fields
are assumed as pulses, with envelopes in the form of Gaussians,
and $T$ is the delay between the two pulsed fields.

The ground state of the system is calculated first without
the application of THz fields. As mentioned, the ground state is
localized on the right-hand-side well of the structure
which has the deepest potential (DW). The numerical method
for this calculation uses imaginary-time propagation
within a split-operator approach~\cite{review, degani}.
The ground state $\psi_0(z)$ is used as the initial state $\Psi(z,0)$
to be advanced in time under the full hamiltonian Eq.~(\ref{H}) by
performing the operation in small time increments $\Delta t$:
\begin{equation}
\Psi ( z,t + \Delta t ) = e^{ -iH\Delta t / \hbar } 
\Psi ( z,t ).
\label{eq1}
\end{equation}
Details of the numerical procedure are given in Refs.~\onlinecite{review, degani}.

With the application of a THz field, which can have the field frequency and intensity varied,
the initial state can be coupled to excited states to create a particle current
flowing to both sides of the structure. 
In the present work the exciting fields are chosen
with energy that is sufficient to excite
current only towards the right side of the system.
The current is calculated at a given point $z=z_c$ in the
(right-hand side) region of the continuum by the expression
\begin{equation}
J(t)=\Re \left( \frac{\hbar}{i m^*} \Psi(z_c,t)^* \frac{\partial
 \Psi(z_c,t)}{\partial z}   \right),
\label{J}
\end{equation}
from which the integrated current is calculated
\begin{equation}
I = \int J(t) dt.
\label{I}
\end{equation}

Since the numerical method employed uses hard-wall boundary conditions,
meaning vanishing wavefunctions at the boundaries of the system, these currents can be reflected
at the system boundaries and produce interference effects. To avoid these effects
in the calculation of the field-induced current, a spatially large system is used
with the inclusion of imaginary-potential barriers on its edges.
These regions act as damping barriers and minimize wavefunction reflections.
The important parameters to be set for these damping barriers are basically
their spatial extent and their potential heights. Limiting values
for these parameters are given in the 
work of Neuhasuer and Baer\cite{neuhasuer} and they depend on the
energy of the electron impinging on the barrier. Since our electron current are
the result of THz-photon absorption, the energy of the electron contributing to the
current depends on the exciting field frequency. This makes the calibration of
these parameters a little tricky, yet it is still possible to considerably reduce
unwanted reflection-effects in this way.

\section{Results and discussion}

The asymmetric double well system investigated is as shown in
Fig.~1, and
consists basically of a one-dimensional problem with
the band-offset energies
taken from the materials GaAs/Al$_x$Ga$_{1-x}$As that
make up real systems.
The effective mass and dielectric constant used are those of GaAs,
namely
$m^* = 0.067m_0$, $m_0$ being the bare electron mass, and 
$\epsilon = 12.5$.
The band-offset energies are given
as a function of the Al concentration $x$
by the following equation,\cite{concentration}
\begin{equation}
V(x) = 0.6 \times ( 1155 x + 370 x^2 ) \;\;\;\;\;[{\rm meV}]
\;\;\;\;.
\label{conc}
\end{equation}

The regions of the structure will henceforth be labeled,
starting on the left side as:
left barrier (LB), shallower well (SW), middle barrier (MB),
deeper well (DW), right barrier (RB), and continuum region to the right (CT).

The thickness of the barriers and wells, inasmuch as
their potential heights (i.e. the Al concentration $x$) were chosen carefully
in order to have a ground state (state-1 in Fig.~1)
localized in the deepest well with the next
two excited states (state-2 and -3) forming almost a doublet, and
a fourth state (state-4) still
bound to the double well. In fact, all these excited states are virtual
bound states, since the presence of the continuum (CT) allows for only one
truly bound state (the ground state). An additional characteristic was desired,
namely, that the energy separation between the doublet of excited states (state-2 and 3)
and the upmost higher energy state (state-4) should be smaller than the
energy difference between the state-1 and the onset of the continuum.
This warrants that exciting fields with frequencies 
capable of coupling states-2 and 3 with state-4 will not create currents
due to transitions from the ground state at DW to the continuum (CT).

A set of system parameters attending the above specifications is found:
\begin{itemize}
\item LB: Al$_{0.4}$Ga$_{0.6}$As, 240 nm thick;
\item SW: Al$_{0.2}$Ga$_{0.8}$As, 9 nm thick;
\item MB: Al$_{0.4}$Ga$_{0.6}$As, 3 nm thick;
\item DW: GaAs, 7.7 nm thick;
\item RB: Al$_{0.4}$Ga$_{0.6}$As, 1.5 nm thick;
\item CT: Al$_{0.2}$Ga$_{0.8}$As, 750 nm thick.
\end{itemize}
To help finding this set of parameters,
RB height was first taken equal to the LB (both with Al concentration
$x$=0.4). In this case the four
states are truly bound states with energies: E$_1$ = 50 meV,  E$_2$ = 180 meV,
E$_3$ = 190 meV and E$_4$ = 270 meV, with respect to the bottom of DW (the GaAs well).
When CT is set with $x$=0.2, a continuum of states is formed in the far right region
and affects these energies. Next we discuss how this can be seen from the
THz-induced current calculations.

To calculate the field-induced current two damping barriers at the system edges
were used. A 30-nm thick damping barrier was defined on the far left side of LB
and similarly a 40-nm thick barrier was placed on the far right side of CT.
Both imaginary barriers had a potential of approximately $-i$50 meV.
The particle current (Eqs.(\ref{J}) and (\ref{I})) were calculated in a position 
$z_c$ just before the damping barrier in CT.

In Fig.~1(b) the integrated current induced by the probe field (switch is off)
is shown as a function of the probe frequency (energy) in a logarithmic scale.
The inset shows the same but in a linear vertical scale.
It is observed that for probe energies less
than $\sim$ 98 meV there is almost zero current in view of the fact
that the THz photon energy is not sufficient to excite the ground state
to at least the onset of the continuum in CT. Above this limiting
energy the current grows up to finite values and
manifests two strong peaks about 127 meV and 143 meV, and
a weaker peak at 225 meV. These peaks in the spectrally resolved current
are related to the transitions induced by the THz field from
the ground state to the (virtual) excited states (states-2, -3, and -4)
followed by tunneling to the continuum (see Fig.~1(b)).
It is interesting to note the similarity of this spectra for the induced
current with the optical spectra for the intersubband transitions calculated
by Wu {\it et al.}\cite{wu}. Like the optical spectra,
the induced current (or photocurrent) spectra
shows two peaks with a strong depletion in between resulting from
interference effects with the nearby continuum, that is, a Fano-like
effect. This strong suppression of current between the two peaks
can be manipulated to create
switching devices with fast responses, 
as pointed out by Wu {\it et al.}\cite{wu} in their
study for the optical case. Here, we explore similar effects
however looking at the photocurrent generated in the process.
Before turning the switch field on, we further discuss
the effects of the probe field on the induced current.

\begin{figure}[htbp]
\centerline{\epsfig{file=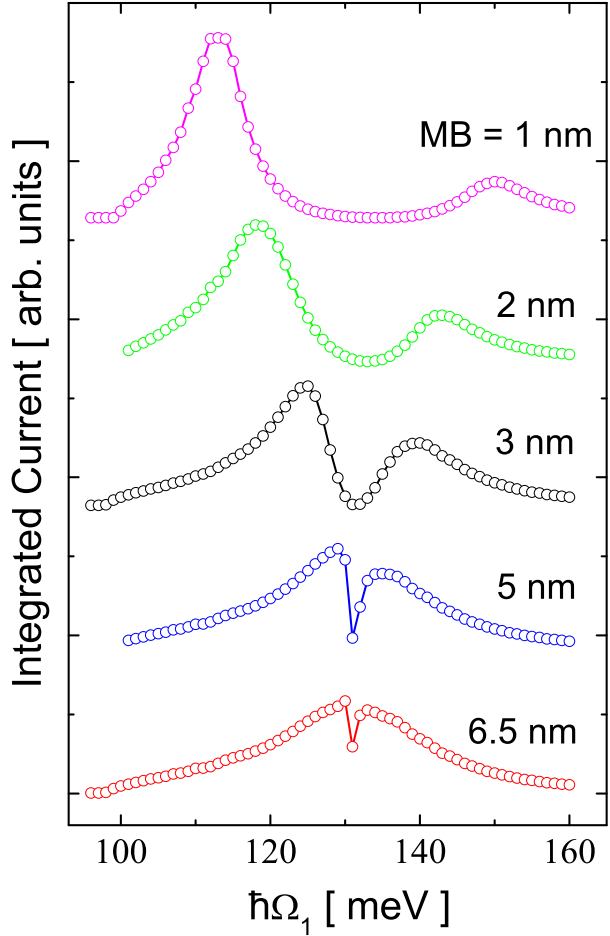,width=8.0cm}}
\caption{(Color online) Field-induced current spectra for systems with different
thickness of the middle barrier MB.  
The thickness of RB is kept at 1.5 nm and
the probe field intensity is $F_1$=10 kV/cm.
The current spectra are offset vertically for clarity.}
\label{B1}
\end{figure}

In Fig.~\ref{B1} the induced current is shown for systems
with different thicknesses of the middle barrier MB, with the spectra
focused on the two stronger peaks related to the transitions
1$\rightarrow$2 and 1$\rightarrow$3. The effect of increasing the
thickness of MB is to decouple the two quantum wells.
As MB becomes wider, the
doublet weakens and the (virtual) excited states become more
independently localized in different quantum wells. 
State-2, mostly localized
in the SW, contributes little to the current since the transition
1$\rightarrow$2 is weakened with  the decrease of both the spatial
overlap between states 1 and 2, and
the probability of tunneling to the continuum.
This is seen in Fig.~\ref{B1} with the merging of the two peaks
in a one (1$\rightarrow$3). In this process, the interference effects
with the continuum are retained.

\begin{figure}[htbp]
\centerline{\epsfig{file=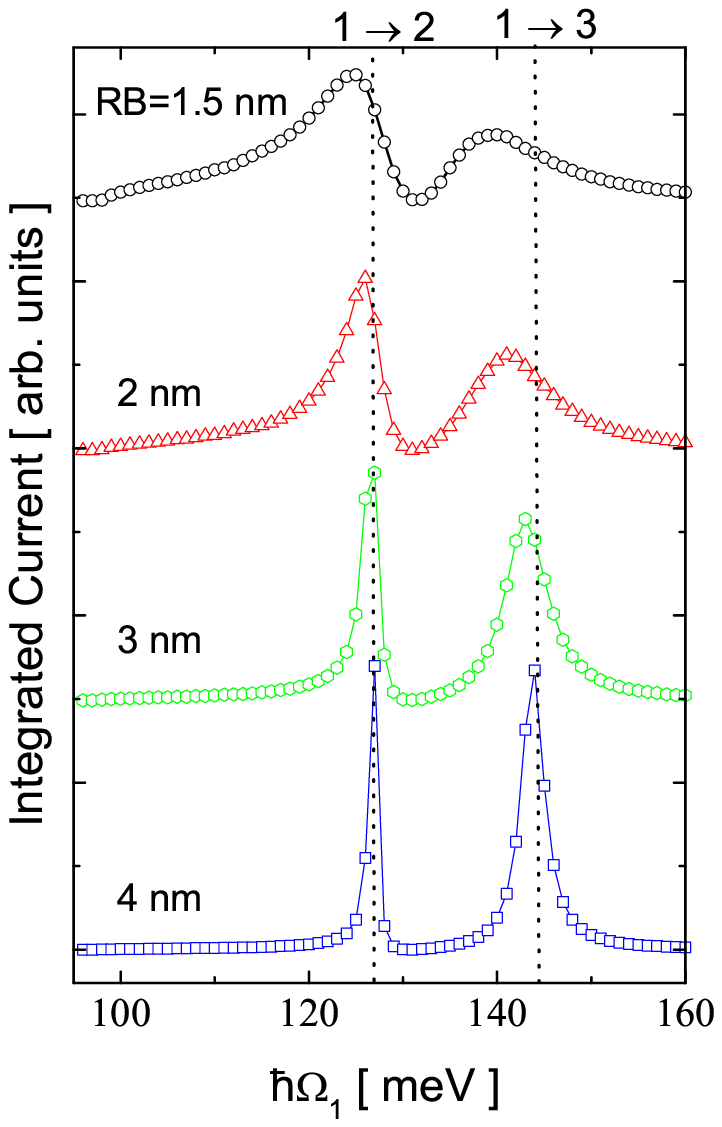,width=8.0cm}}
\caption{(Color online)
 Field-induced current spectra for systems with different
thicknesses of the right barrier RB.  
Thickness of middle barrier MB is kept at 3 nm
and the probe field intensity is $F_1$=10 kV/cm
while the switch field $F_2=0$.
Vertical dotted lines indicate the energy transitions for a system
without the continuum (RB thickness infinite).
Current spectra split vertically for better visualization.   }
\label{B2}
\end{figure}

In Fig.~\ref{B2} the coupling with the continuum is investigated by 
varying the thickness of the right barrier RB. As RB
becomes wider,
the states in the double well are less affected by the continuum.
It is then observed that the peaks become sharper, indicating 
longer lifetimes (lesser couplings to the continuum), and it is also observed a large
splitting between the peaks, since the states effectively become more localized
in the double well region. Note that for large RB thicknesses
(near 4 nm in Fig.~\ref{B2}),
the energies of the two peaks tend to the values expected for the double well
system in the absence of the continuum.

\begin{figure}[htbp]
\centerline{\epsfig{file=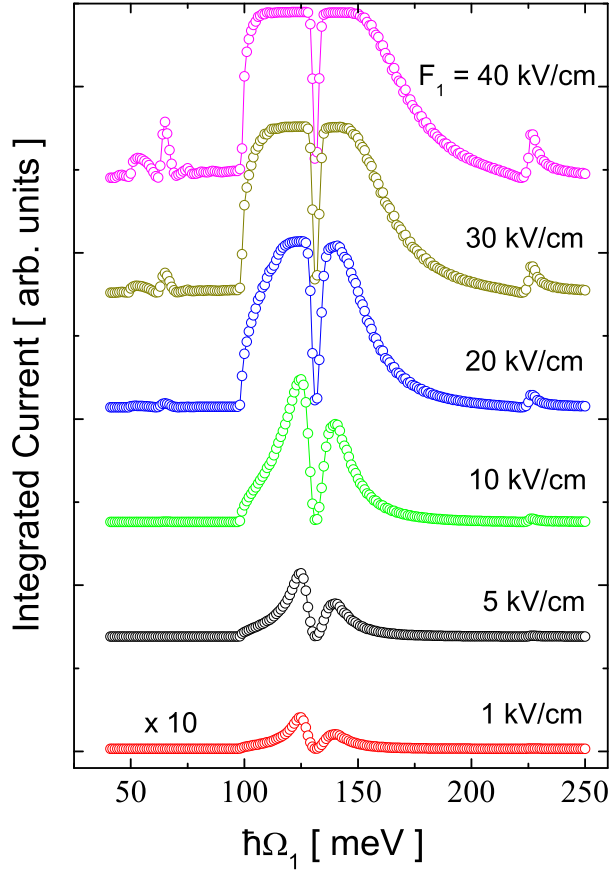,width=8.0cm}}
\caption{(Color online)
 Field-induced current spectra for different
intensities of the exciting probe field.  
Thickness of MB is 3 nm and of RB is 1.5 nm.
Current spectra split vertically for better visualization.   }
\label{F1}
\end{figure}

Next, the intensity of the probe field is varied, with
the result shown in Fig.~\ref{F1}.
For moderate intensities
$<$10 kV/cm the two peaks for the transitions 1$\rightarrow$2 and 1$\rightarrow$3
dominate the spectrum. For stronger intensities the peak for the transition
1$\rightarrow$4, at the higher energy side of the spectrum, has its
amplitude increasing 
linearly with the field intensity.
For the lower energy side of the spectrum, a double peak appears whose
amplitude increases
superlinearly with the field's intensity. This part of the spectrum is not allowed
for one-photon transitions due to energy conservation. Hence, this peak
is attributed to a two-photon process for the transitions
1$\rightarrow$2 and 1$\rightarrow$3. The superlinear growth of this double peak
and its energy in the spectrum (half of that permitted
for the one-photon process) strongly supports this interpretation.
Another feature seen in Fig.~\ref{F1} is a saturation effect for energies
around the 1$\rightarrow$2 and 1$\rightarrow$3 transitions.
The relatively small values for the intensities of the fields
that produce
such a saturation are spurious and
follows from the single-electron physics used to
calculate the current. For larger fields,
and strong transition rates, a single-electron
is completely ionized by the THz field, causing
the integrated current to saturate.
For an actual system with electrons provided by dopant impurities, the saturation
(complete ionization) is expected to occur at higher field intensities.

\begin{figure}[htbp]
\centerline{\epsfig{file=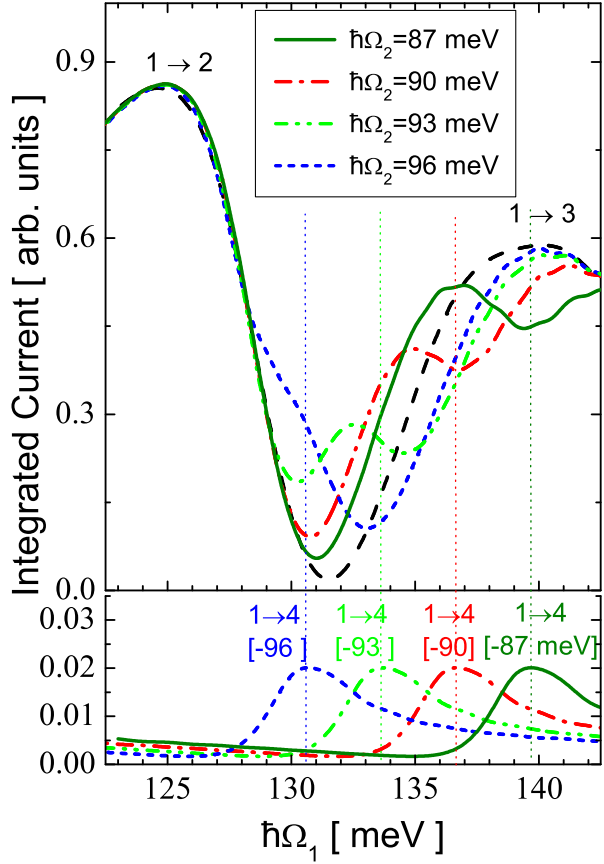,width=8.0cm}}
\caption{(Color online)
 Upper panel: Field-induced current spectra for different
frequencies of the switch field.  
Thickness of MB is 3 nm and of RB is 1.5 nm; probe and
switch fields'
intensities are F$_1$ = 10 kV/cm and F$_2$=20 kV/cm,
respectively.
Dashed line: F$_2$ = 0.
Lower panel: Part of the spectra showing the transition 1$\rightarrow$4
redshifted by the switch photon energy.}
\label{E2}
\end{figure}

When a second (switch) field is used, the spectrum is modified
especially in the region between the transitions 1$\rightarrow$2 and 1$\rightarrow$3 
where interference with the continuum strongly supresses the
induced current.
This is due to the absorption of an additional photon that excites
the transition from states 2 and 3 to the higher energy state-4.
Figure~\ref{E2} shows the spectra for a system with barriers' thicknesses
of 3 nm and 1.5 nm for MB and RB, respectively, for a
probe intensity F$_1$= 10 kV/cm and a switch field
intensity F$_2$ = 20 kV/cm. The two pulses have the same
time spams (2 ps) and phases and have zero delay between them
[$\phi$=0 and $T$=0 in Eq.~(\ref{H})].
The frequency of the switch is varied slightly
around the energy difference between the states-(2,3) and state-4.
Note that the switch field alone does not induce current for these
frequencies since the corresponding photon energy is
 smaller than the energy difference between the
ground state and the onset of the continuum, which is $\approx$ 98 meV.
The system parameters were carefully chosen to yield such a behavior.

In the lower panel of Fig.~\ref{E2} the peak corresponding to transition
1$\rightarrow$4 is seen to be redshifted by exactly the
switch field energy $\hbar \Omega_2$.
It shows that the part of the spectrum (upper panel) that is affected by the switch
is the one in the energy range that matches an additional transition
to the state-4 by the absorption of a switch photon.
The effect of the switch can be either weakening an existing current by
further excitation to the state-4 (as in Fig.~\ref{E2}
for $\hbar \Omega_2$= 87 meV) or enhancing an otherwise
suppressed current (as it is the case
for $\hbar \Omega_2$= 93 meV) by modifying the interference with
the continuum. This latter case was explored by Wu {\it et al.}\cite{wu} 
as a possible all-optical switching mechanism, since the switch field
strongly modifies the optical absorption of the probe field.
We observe similar effects, not on the optical absorption, but instead on the
induced current, meaning that the optical switching mechanism can have a similar
signature on its photocurrent. It is observed, for a probe field
with frequency set at the zero current ($\hbar \Omega_1 \approx$ 132 meV)
and a switch field with $\hbar \Omega_2 \approx$ 93 meV, that neither
of the two fields alone suffices to
induce current, yet when they are simultaneously present there is 
an appreciable current induced (that is, a current that is larger
than that originated in two-photon processes).
This pattern can be used in a logic gate integrating optics and electronics.

\begin{figure}[htbp]
\centerline{\epsfig{file=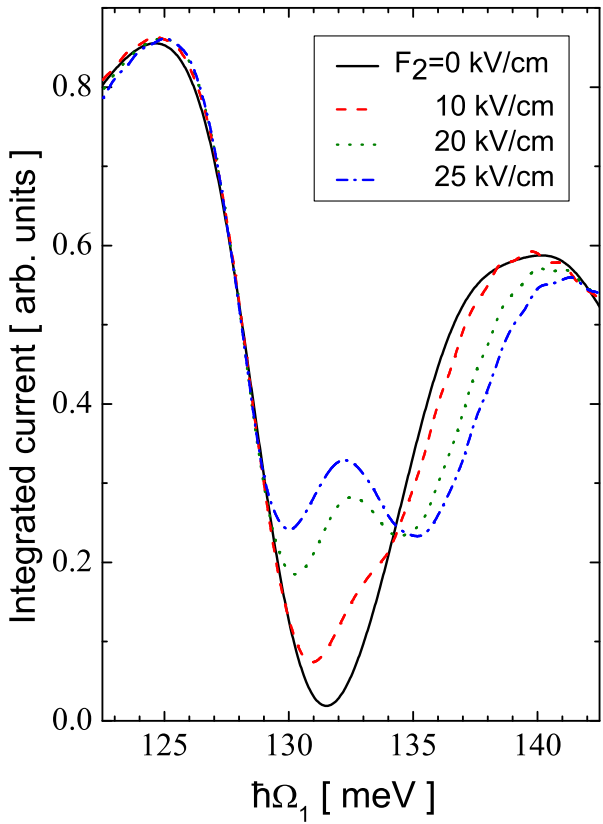,width=8.0cm}}
\caption{(Color online)
 Field-induced current spectra for different
intensities of the switch field. 
Thickness of MB is 3 nm and of RB is 1.5 nm; probe field
intensity F$_1$ = 10 kV/cm, and switch field frequency
$\hbar \Omega_2$= 93 meV.    }
\label{F2}
\end{figure}

Finally, in Fig.~\ref{F2} the effect of the switch-field's intensity on
the induced current is shown for a probe intensity F$_1$ = 10 kV/cm.
As it would be expected, increasing the switch intensity enhances the
switching effects just described.
We have obtained results similar to the ones shown in Fig.~\ref{F2}
by increasing the delay time between the two pulses [$T$ in Eq.~(\ref{H})] ,
that is, the delay effectively lowers the intensity of the switch field
that is necessary to produce the desired switching effect.
No changes were observed in the integrated current when varying
the relative phase $\phi$ in  Eq.~(\ref{H}).

In conclusion, the photocurrent induced by THz fields was investigated theoretically
for a asymmetric double quantum well structure. The system parameters were
chosen to have two excited (virtual) states forming a doublet which interacts 
with a nearby continuum yielding an almost zero induced current for probe field frequencies
between the energies of the states in the doublet. A second (switch)
field was used to modify the interference with the continuum by
coupling the doublet to a higher excited state. The resulting
current could be used as a logic gate integrating optics and electronics.
Besides this two-field (probe/switch) investigation, the induced current was studied
with varying system parameters, and also for nonlinear effects, such as the two-photon
absorption for intense fields. 
Although the study contemplated a simple one-dimensional model,
similar trends should be expected for electrons bound, for example, in
a quantum dot and interacting with a nearby a continuum, such as a wetting layer.
The THz fields were assumed as pulses and the initial state
evolved in time was a single electron only for convenience
in the numerical calculations. More realistic conditions in actual
situations where such logic mechanism would operate, for example in 
continuous excitation, with the presence
of many electrons, or under large bandwidth field excitation may enrich the problem
but the trends found here can be the starting point for future investigations
of the induced photocurrent.

\section*{ACKNOWLEDGMENT}

MZM, MHD, and PFF gratefully acknowledge financial support from
Programa Institutos Nacionais de Ci\^ encia e
Tecnologia and Conselho Nacional de Desenvolvimento Cient\' \i fico
e Tecnol\' ogico - CNPq/MCT.
MHD thankfully acknowledges financial support from Dhiminsky ATA.

\end{document}